\documentclass[12pt]{article}
\usepackage{bm}
\renewcommand{\d}{\partial}
\newcommand{\LQCD}{\Lambda_{\rm QCD}}
\newcommand{\LambdaQCD}{\Lambda_{\rm QCD}}
\def\U1A{U(1)$_{\rm A}$}
\newcommand{\<}{\langle}
\renewcommand{\>}{\rangle}
\newcommand{\+}{\dagger}

\def\beq{\begin{equation}}
\def\eeq#1{\label{#1}\end{equation}}
\def\eeqn{\end{equation}}

\def\beqa{\begin{eqnarray}}
\def\eeqa#1{\label{#1}\end{eqnarray}}
\def\eeqan{\end{eqnarray}}

\let\bar=\overbar

\def\Dslash{\not{\hbox{\kern-4pt $D$}}}
\def\dslash{\not{\hbox{\kern-2pt $\del$}}}

\def\msb{{\bar{\ssstyle M \kern -1pt S}}}

\def\Title#1{\begin{center} {\Large {\bf #1} } \end{center}}

\begin{document}

\Title{Confinement and domain walls in high density quark matter}

\bigskip\bigskip


\begin{raggedright}  

{\it Dam T. Son\\
Physics Department\\
Columbia University\\
New York, NY 10027, USA}
\bigskip\bigskip
\end{raggedright}

In this talk I will review some recent progress in our understanding
of properties of high density quark matter.  This talk is based mainly
on the papers~\cite{deconf}--\cite{bec}, done in collaboration with
D.~Rischke, M.~Stephanov, and A.~Zhitnitsky.  

In recent years, our knowledge of dense quark matter has considerably
expanded.  We now understand that, quark matter at high densities
exhibits the phenomenon of color superconductivity, which determines
the symmetry of the ground state and its infrared dynamics.  In the
simplest case of $N_f=2$ massless quarks, the ground state of at high
baryon densities is the 2SC state \cite{2SC}, characterized by the
condensation of diquark Cooper pairs.  These pairs are antisymmetric
in spin ($\alpha,\beta$), flavor ($i,j$) and color ($a,b$):
\begin{eqnarray}
  \<q^{ia}_{L\alpha} q^{jb}_{L\beta} \>^* &=&  
  \epsilon_{\alpha\beta} \epsilon^{ij}\epsilon^{abc} X^c \, ,
  \nonumber \\
  \<q^{ia}_{R\alpha} q^{jb}_{R\beta} \>^* &=&  
  \epsilon_{\alpha\beta} \epsilon^{ij}\epsilon^{abc}Y^c \, .
  \label{2flavor}
\end{eqnarray}
$X^c$ and $Y^c$ are complex color 3-vectors.  In the ground state,
they align along the same direction in the color space and break the
color SU(3)$_c$ group down to SU(2)$_c$.  Thus, five of the original
eight gluons acquire ``masses'' by the Meissner effect, similar to the
Higgs mechanism.  The remaining three gluons are massless
(perturbatively).  Because of Cooper pairing, the spectrum of quark
excitations carrying SU(2) color charge has a gap $\Delta$.

The physics below the energy scale $\Delta$ is governed by the pure
gluodynamics in the remaining unbroken SU(2) sector.  As shown in
Ref.\ \cite{deconf}, the process of high-density ``deconfinement'' is
quite nontrivial in this case: the quarks are {\em always} confined,
however, the confinement radius grows {\em exponentially} with
increasing density.

Below the scale $\Delta$, heavy (gapped) degrees of freedom decouple
and the remaining fields can be described by a local effective
Lagrangian.  The absence of quarks carrying SU(2) charges below
$\Delta$ (all are bound into SU(2) singlet Cooper pairs) implies that
the medium is transparent to the SU(2) gluons: there is no Debye
screening and Meissner effect for these gluons.  Mathematically, the
gluon polarization tensor $\Pi^{\mu\nu}_{ab}(q)$ vanishes at $q=0$,
which can be checked by a direct calculation of $\Pi$ at small $q$
\cite{Rischke2fl}.  However, although a static SU(2) charge cannot be
completely Debye screened by SU(2) neutral Cooper pairs, it can still
be {\em partially} screened if the medium is polarizable, i.e., if it
has a {\em dielectric constant} $\epsilon$ different from unity.
Analogously, the medium can, in principle, have a magnetic
permeability $\lambda\neq1$.

The requirements of locality, gauge and rotational invariance, and
parity fixes the effective action below the scale $\Delta$ in the
following form
\begin{equation}
  S_{\rm eff} = {1\over g^2}\int\!d^4x\,\biggl({\epsilon \over 2}
  {\bf E}^a\cdot {\bf E}^a - {1\over2\lambda}{\bf B}^a\cdot {\bf B}^a
  \biggr) \, ,
  \label{Leff}
\end{equation}
where $E^a_i\equiv F_{0i}^a$ and
$B^a_i\equiv{1\over2}\epsilon_{ijk}F^a_{jk}$.  Explicit calculations
\cite{deconf} gives
\begin{eqnarray}
  \epsilon &=& 1 + \kappa = 1 + {g^2\mu^2\over18\pi^2\Delta^2}
    \, , \label{epsilon} \\
  \lambda &=& 1 \, . \label{mu}
\end{eqnarray}
Since at high densities, the gap $\Delta$ is exponentially suppressed
compared to the chemical potential $\mu$ \cite{Son}, $\epsilon\gg1$,
i.e., the dielectric constant of the medium is very large.  This can
be interpreted as a consequence of the fact that the Cooper pairs have
large size (of order $1/\Delta$) and so are easy to polarize.  The
magnetic permeability, in contrast, remains close to 1 due to the
absence of mechanisms that would strongly screen the magnetic field.

The effective strong coupling constant $\alpha_s'$ in the theory 
(\ref{Leff}), the
equivalence of $\alpha_s=g^2/(4\pi\hbar c)$ in the vacuum, is very
small at the matching scale with the microscopic theory (i.e.,
$\Delta$).  In our dielectric medium, the Coulomb potential between
two static charges separated by $r$ is $g^2/(\epsilon r)$.  Thus, we
have to replace $g^2$ by $g^2_{\rm eff}=g^2/\epsilon$.  The velocity
of light $c$ also needs to be replaced by the velocity of gluons
$v=1/\sqrt{\epsilon}$.  This gives
\begin{equation}
  \alpha_s' =
  {g_{\rm eff}^2\over 4\pi \hbar v} = {g^2\over4\pi\sqrt{\epsilon}} 
  = {3\over 2\sqrt{2}} {g\Delta\over\mu}\, .
  \label{alphaeff}
\end{equation}
The coupling increases logarithmically as one moves to lower energies,
since pure SU(2) Yang-Mills theory is asymptotically free.  This
coupling becomes large at the confinement scale $\LQCD'$, which is the
mass scale of SU(2) glueballs.  Since $\alpha_s'$ is tiny (because
$\Delta/\mu\ll1$), it takes long to grow, and the scale $\LQCD'$ is
thus very small.  Using the one-loop beta function, one can estimate
\begin{equation}
  \LQCD' \sim \Delta \exp\biggl(-{2\pi\over\beta_0\alpha_s'}\biggr)
  \sim \Delta \exp\biggl(-{2\sqrt{2}\pi\over11}{\mu\over g\Delta}
  \biggr) \, ,
  \label{LQCD'}
\end{equation}
where $\beta_0=22/3$ in SU(2) gluodynamics.  The possible relevance of
the light glueballs for astrophysics was discussed by Ouyed and
Sannino \cite{Ouyed}.

In Ref.\ \cite{domain} we show that at high baryon densities QCD must
have domain walls.  The simplest case allowing for domain walls is
again $N_f=2$ massless flavors.  In perturbation theory, there is a
degeneracy of the ground state with respect to the relative U(1) phase
between $X^a$ and $Y^a$ in Eq.\ (\ref{2flavor}).  This is due to the
\U1A symmetry of the QCD Lagrangian at the classical level.  This fact
implies that the \U1A symmetry is spontaneously broken by the
color-superconducting condensate.  Since this is a global symmetry,
its breaking gives rise to a Goldstone boson, which carries the same
quantum numbers as the $\eta$ boson in vacuum.

The field corresponding to $\eta$ boson can be constructed explicitly.
Indeed
\begin{equation}
  \Sigma=XY^\+\equiv X^aY^{a*} \, ,
  \label{SigmaXY}
\end{equation}
in contrast to $X$ and $Y$, is a gauge-invariant order parameter.
Furthermore $\Sigma$ carries a nonzero \U1A charge.  Under the \U1A
rotations
\begin{equation}
  q\to e^{i\gamma_5\alpha/2}q \, ,
  \label{qalpha}
\end{equation}
the fields (\ref{2flavor}) transform as $X\to e^{-i\alpha}X$, $Y\to
e^{i\alpha}Y$, and therefore $\Sigma\to e^{-2i\alpha}\Sigma$.  Thus,
the color-superconducting ground state, in which $\<\Sigma\>\neq0$,
breaks the \U1A symmetry.  The Goldstone mode $\eta$ of this symmetry
breaking is described by the phase $\varphi$ of $\Sigma$,
\begin{equation}
  \Sigma=|\Sigma|e^{-i\varphi} \, . 
\end{equation}
Under the \U1A rotation (\ref{qalpha}), $\varphi$ transforms as
\begin{equation}
  \varphi \to \varphi + 2\alpha \, .
  \label{phi_transf}
\end{equation}

At low energies, the dynamics of the Goldstone mode $\varphi$ is
described by an effective Lagrangian, which must take the following
form,
\begin{equation}
  L = f^2 [(\d_0\varphi)^2 - u^2 (\d_i\varphi)^2] \, .
  \label{Leffnomass}
\end{equation}
Two free parameters of this Lagrangian are the decay constant $f$ of
the $\eta$ boson, and its velocity $u$.  In general, $u$ may be
different from 1 since the Lorentz invariance is violated by the dense
medium. For large chemical potentials $\mu\gg\LambdaQCD$, the leading
perturbative values for $f$ and $u$ have been determined by
Beane {\em et al.} \cite{BBS}:
\begin{equation}
  f^2 = {\mu^2\over 8\pi^2} \, , \qquad  u^2 = {1\over3} \, .
\end{equation}
In particular, the velocity of the $\eta$ bosons, to this order, is
equal to the speed of sound.  The fact that $f\sim\mu$ plays an
important role in our further discussion.

It is well known that the \U1A symmetry is not a true symmetry of the
quantum theory, even when quarks are massless.  The violation of the
\U1A symmetry is due to non-perturbative effects of instantons.  Since
at large chemical potentials the instanton density is suppressed (see
below), the $\eta$ boson still exists but acquires a finite mass.  In
other words, the anomaly adds a potential energy term $V_{\rm
inst}(\varphi)$ to the Lagrangian (\ref{Leffnomass}),
\begin{equation}
  L = f^2 [(\d_0\varphi)^2 - u^2 (\d_i\varphi)^2] -
      V_{\rm inst}(\varphi) \, .
  \label{LeffV}
\end{equation}
The curvature of $V_{\rm inst}$ around $\varphi=0$ determines the mass
of the $\eta$.

A standard symmetry argument determines periodicity of $V_{\rm
inst}(\varphi)$.  One can formally restore the \U1A symmetry by
accompanying (\ref{qalpha}) by a rotation of the $\theta$-parameter
\begin{equation}
  \theta \to \theta + N_f \alpha = \theta + 2\alpha \, .
  \label{theta_transf}
\end{equation}
This symmetry must be preserved in the effective Lagrangian, so the
latter is invariant under (\ref{phi_transf}) and (\ref{theta_transf}).
This means that the potential $V_{\rm inst}$ is a function of the
variable $\varphi-\theta$, unchanged under \U1A.  Since we know that
the physics is periodic in $\theta$ with period $2\pi$, we can
conclude that, at the physical value of the theta angle $\theta=0$,
$V_{\rm inst}$ is a periodic function of $\varphi$ with period $2\pi$.

Moreover, at large $\mu$, $V_{\rm inst}$ can be found from instanton
calculations explicitly.  The infrared problem that plagues these
calculations in vacuum disappears at large $\mu$: large instantons are
suppressed due to Debye screening.  As a result, most instantons have
small size $\rho\sim{\cal O}(\mu^{-1})$ and the dilute instanton gas
approximation becomes reliable. One-instanton contribution,
proportional to $\cos(\varphi-\theta)$, dominates in $V_{\rm
inst}$. Therefore,
\begin{equation}
  V_{\rm inst}(\varphi) = -a \mu^2\Delta^2\cos\varphi \, ,
  \label{Vinst}
\end{equation}
where $\Delta$ is the BCS gap, and $a$ is a dimensionless function of
$\mu$ estimated in Ref.\ \cite{domain}.  Here we only note that $a$
vanishes in the limit $\mu\to\infty$.  This is an important fact,
since it implies that the mass of the $\eta$ boson,
\begin{equation}
   m = \sqrt{a\over2} \, {\mu\over f} \Delta 
   = 2\pi \sqrt a \Delta \, ,
   \label{meta}
\end{equation}
becomes much smaller than the gap $\Delta$ at large $\mu$.  In this
case the effective theory (\ref{LeffV}) is reliable, since meson modes
other than $\eta$ have energy of order $\Delta$, i.e., are much
heavier than $\eta$ and decouple from the dynamics of the latter.

The Lagrangian (\ref{LeffV}) with the potential (\ref{Vinst}) is just
the sine-Gordon model, in which there exist domain-wall solutions to
the classical equations of motion.  The profile of the wall parallel
to the $xy$ plane is
\begin{equation}
   \varphi = 4 \arctan e^{mz/u} \, ,
\end{equation}
so the wall interpolates between $\varphi=0$ at $z=-\infty$ and
$\varphi=2\pi$ at $z=\infty$.  The tension of the domain wall is
\begin{equation}
  \sigma = 8\sqrt{2a}\, uf\mu\Delta  
  \, .
  \label{sigma}
\end{equation}
A good analog of this domain wall is the $N=1$ axion domain wall,
which also interpolates between the same vacuum.

Another phase with domain walls is that of kaon condensation.  The
likelihood of this phase was recently emphasized by Sch\"afer and 
other authors
\cite{kaoncond}.  The crucial observation is that kaons have small
masses \cite{inverse} in the color-flavor locked phase (CFL)
\cite{CFL} of high-density QCD.  A relatively small strangeness
chemical potential is thus sufficient to drive kaon condensation.
Moreover, it was argued that the mass of the strange quark also works
in favor of kaon condensation. In contrast to the conventional charge
kaon condensation in nuclear matter, in the CFL phase it is the
neutral kaons which are more likely to condense.  This is due to the
inverse mass ordering \cite{inverse} of mesons in the CFL phase, which
makes the neutral kaons lighter than the charge kaons, at least at
very high densities.

In a recent paper \cite{kaon} we show that the $K^0$-condensed phase
has in its spectrum an extremely light bosonic particle, whose
presence implies the existence of non-topological metastable domain
walls.  Again, this feature can be understood easily from symmetry
arguments.  The $K^0$ condensate spontaneously breaks a global U(1)
symmetry: the strangeness.  The phase of this condensate becomes a
Goldstone boson.  However, since strangeness is violated by weak
processes, the would-be Goldstone boson acquires a very small mass
proportional to $G_F^{1/2}$.  At very low energies, the effective
Lagrangian for $\varphi$ must have the form of the sine-Gordon theory,
which possesses domain wall solutions interpolating between
$\varphi=0$ and $\varphi=2\pi$.  More recently, vortons were discussed
by Kaplan and Reddy \cite{vorton}.

It is interesting to note that if baryon number is violated, the
superfluid Goldstone mode also acquires a mass.  Since the superfluid
order parameter is a dibaryon, the mass square of the superfluid
Goldstone boson is proportional to the amplitudes of the $\Delta B=2$
transitions.  The experimental bound on $n\bar n$ oscillations,
$\tau_{n\leftrightarrow\bar n}>10^8$ s, put an upper limit on the mass
of the Goldstone boson, $m<10^{-7}$ eV.  The thickness of the
corresponding domain wall is larger than about 1 m, and still might be
less than the radius of neutron stars.  However, unless the neutron
star under consideration rotates very slowly, a domain wall that thick
is unlikely to exist because of the high density of vortices, which
are separated by distances of order $10^{-2}$ cm.

Finally, domain walls of the type discussed above also exist in
two-component Bose-Einstein condensates (BEC) \cite{bec}.  Recently,
two different hyperfine spin states of $^{87}$Rb, which were condensed
in the same trap by the technique of sympathetic cooling
\cite{Myatt_et_al}.  A similar state has been observed for sodium gas
\cite{Ketterle}.  Binary BEC breaks spontaneously {\em two} global
U(1) symmetries.  It is moreover possible to couple two condensates by
a driving electromagnetic field tuned to the transition frequency.  In
this case atoms can be interconverted between the two spin states and
the numbers of atoms of each species are not conserved separately
anymore; only the total number of atoms is constant.  In this case,
only one U(1) symmetry remains exact, the other one is explicitly
violated.  The violated U(1) group corresponds to the relative phase
between the two condensate.  The Goldstone boson arising from the
spontaneous breaking of this U(1) symmetry acquires a gap and gives
rise to the domain walls.

\bigskip
I am grateful to the R.~Ouyed and F.~Sannino for organizing this
meeting, and thank D.~Rischke, M.~Stephanov and A.~Zhitnitsky for
fruitful collaboration.

\end{document}